\documentclass{iopart}
\usepackage{graphicx}
\usepackage[colorlinks=true,citecolor=blue,urlcolor=blue]{hyperref}
\usepackage{xcolor}
\usepackage{orcidlink}
\usepackage{siunitx}

\usepackage[mathb]{mathabx} 

\newcommand{\chirpmass}{\mathcal{M}_{\rm c}}
\newcommand{\Msun}{M_{\odot}}

\begin{document}
\begin{flushleft}
{\rm RESCEU-20/26}
\end{flushleft}

\title[KAGRA upgrade for multimessenger observations of BNS]{Evaluating KAGRA upgrade scenarios for multimessenger observations of binary neutron stars}

\author{Yuta~Michimura$^{1,2,\star}$\orcidlink{0000-0002-2218-4002}, Soichiro~Morisaki$^{3}$\orcidlink{0000-0002-8445-6747}, Kenta~Hotokezaka$^{1}$, Masaomi~Tanaka$^{4,5}$}
\address{$^1$ Research Center for the Early Universe (RESCEU), Graduate School of Science, University of Tokyo, Bunkyo, Tokyo 113-0033, Japan}
\address{$^2$ Kavli Institute for the Physics and Mathematics of the Universe (Kavli IPMU), WPI, UTIAS, University of Tokyo, Kashiwa, Chiba 277-8568, Japan}
\address{$^3$ Institute for Cosmic Ray Research, University of Tokyo, Kashiwa, Chiba 277-8582, Japan}
\address{$^4$ Astronomical Institute, Tohoku University, Sendai, Miyagi 980-8578, Japan}
\address{$^5$ Division for the Establishment of Frontier Sciences, Organization for Advanced Studies, Tohoku University, Sendai, Miyagi 980-8577, Japan}

\ead{$^{\star}$michimura@resceu.s.u-tokyo.ac.jp}
\vspace{10pt}
\begin{indented}
\item[]\today
\end{indented}

\begin{abstract}
Binary neutron star mergers are key targets for multimessenger astronomy, motivating future upgrades of gravitational-wave detectors. For KAGRA, both broadband sensitivity improvements that increase the binary neutron star detection range, and high-frequency optimizations targeting neutron-star physics are under consideration. We present a computationally efficient framework to evaluate the multimessenger performance of detector upgrades by combining Fisher-matrix estimates of localization area and localization volume with detector duty factors and binary neutron star merger rates. We apply this framework to proposed KAGRA upgrade scenarios within the LIGO--Virgo--KAGRA network. For identical sources, the high-frequency upgrade improves sky localization by about 20\% compared with the broadband option. However, when detection rates are taken into account, the broadband upgrade yields a larger number of well-localized events. Despite its shorter binary neutron star range than the other detectors, the inclusion of KAGRA increases the number of events localized within $10^3~{\rm Mpc}^3$ volume by about 60\%. These results provide a quantitative framework for evaluating future detector upgrades from the perspective of multimessenger observations.
\end{abstract}

\maketitle

\section{Introduction} \label{sec:Introduction}
Since the first detection of gravitational waves in 2015~\cite{GW150914}, more than a decade has passed and over 300 gravitational-wave events have been reported~\cite{GWTC-5.0}, providing an unprecedented view of the Universe through a new observational window. Among these discoveries, the binary neutron star merger event GW170817 stands out as a landmark observation~\cite{GW170817,GW170817Multimessenger}, which demonstrated the potential of multimessenger gravitational-wave astronomy.

The detection by the two LIGO detectors in the United States~\cite{aLIGO} together with the Virgo detector in Italy~\cite{AdV} enabled the source to be localized to an area of approximately 30 deg$^2$, leading to the identification of electromagnetic counterparts across the spectrum, from gamma rays to radio waves. These multimessenger observations provided crucial insights into the origin of short gamma-ray bursts~\cite{GRB170817A}, the nature of kilonova emission and the production of heavy elements through the r-process~\cite{Pian2017,Kasen2017,Tanaka2017}, and the equation of state of neutron stars~\cite{GW170817NSEOS}. They also enabled new tests of gravity and cosmology, including constraints on the propagation speed of gravitational waves~\cite{GRB170817A} and measurements of the Hubble constant~\cite{GW170817H0}. The scientific promise of multimessenger studies has been further highlighted by several astrophysically intriguing merger candidates discovered in recent years, including GRB211211A and GRB230307A, which have provided new insights into the diversity of compact object mergers and their electromagnetic counterparts~\cite{Rastinejad2022,Levan2024}.

However, up to the fourth observing run (O4), only two binary neutron star mergers, GW170817 and GW190425~\cite{GW190425}, have been confidently detected, with multimessenger observations achieved only for GW170817. This is due not only to the relatively low binary neutron star merger rate~\cite{GWTC-5population} but also to the fact that accurate sky localization typically requires observations by a network of detectors with sufficient sensitivity and wide geographical separation. It has also been shown, in the context of the KAGRA detector in Japan~\cite{AsoKAGRA,PTEP01KAGRA}, that even a detector with relatively low sensitivity can make a significant contribution to sky localization when operating as part of the global network~\cite{Alvin2026}, emphasizing that multimessenger capability is determined by the network as a whole rather than individual detector sensitivity alone.

Beyond sky localization, the localization volume is an important figure of merit for multimessenger follow-up, as it determines the number of candidate host galaxies and thus the feasibility of counterpart identification. In this context, the addition of KAGRA can further improve parameter estimation by breaking the degeneracy between luminosity distance and binary inclination through polarization measurements~\cite{Hakeda2018}. This leads to improved distance estimation and consequently a reduction in the localization volume.

During the upcoming fifth observing run (O5), the increased detector sensitivity is expected to further enhance the number of binary neutron star detections and improve localization performance~\cite{ObservingScenarioPaper,Kiendrebeogo2023}. Looking beyond O5, upgrades are being planned across the global gravitational wave detector network to enhance its scientific capabilities in the 2030s. The LIGO and Virgo detectors are expected to undergo further sensitivity improvements beyond their current upgrade programs. In parallel, several upgrade options are under consideration for KAGRA, including broadband sensitivity improvements that increase the binary neutron star range and high-frequency optimizations targeting neutron-star physics~\cite{KAGRAplus,KAGRA10year,KAGRA10yearHW,KAGRA10yearScience}.

Assessing the impact of such detector upgrades on multimessenger observations requires not only estimating detector sensitivity, but also quantifying how often binary neutron star mergers can be localized well enough for electromagnetic follow-up observations. In this paper, we evaluate future KAGRA upgrade scenarios from the perspective of multimessenger observations. To enable such a comparison, we develop a computationally efficient framework based on Fisher-matrix estimates of sky localization area and luminosity distance uncertainties in the LIGO--Virgo--KAGRA network, combined with binary neutron star detection rates and detector duty factors. Since a significant fraction of events are observed when one or more detectors are unavailable, incorporating realistic duty factors is important for assessing the performance of detector networks. Using this framework, we compare broadband and high-frequency upgrade options in terms of their ability to increase the number of well-localized events available for electromagnetic follow-up observations.

The remainder of this paper is organized as follows. In Sec.~\ref{sec:Upgrade}, we summarize the detector sensitivity curves for the KAGRA upgrade scenarios considered in this study. In Sec.~\ref{sec:Framework}, we present the framework for estimating 2D sky localization and 3D localization volume in the LIGO--Virgo--KAGRA network. In Sec.~\ref{sec:Results}, we present the results comparing broadband and high-frequency upgrade options in terms of their impact on multimessenger observations. Finally, Sec.~\ref{sec:Conclusions} summarizes our conclusions.

\section{KAGRA upgrade scenarios} \label{sec:Upgrade}

\begin{figure}[t]
	\centering
    \includegraphics[width=0.95\linewidth]{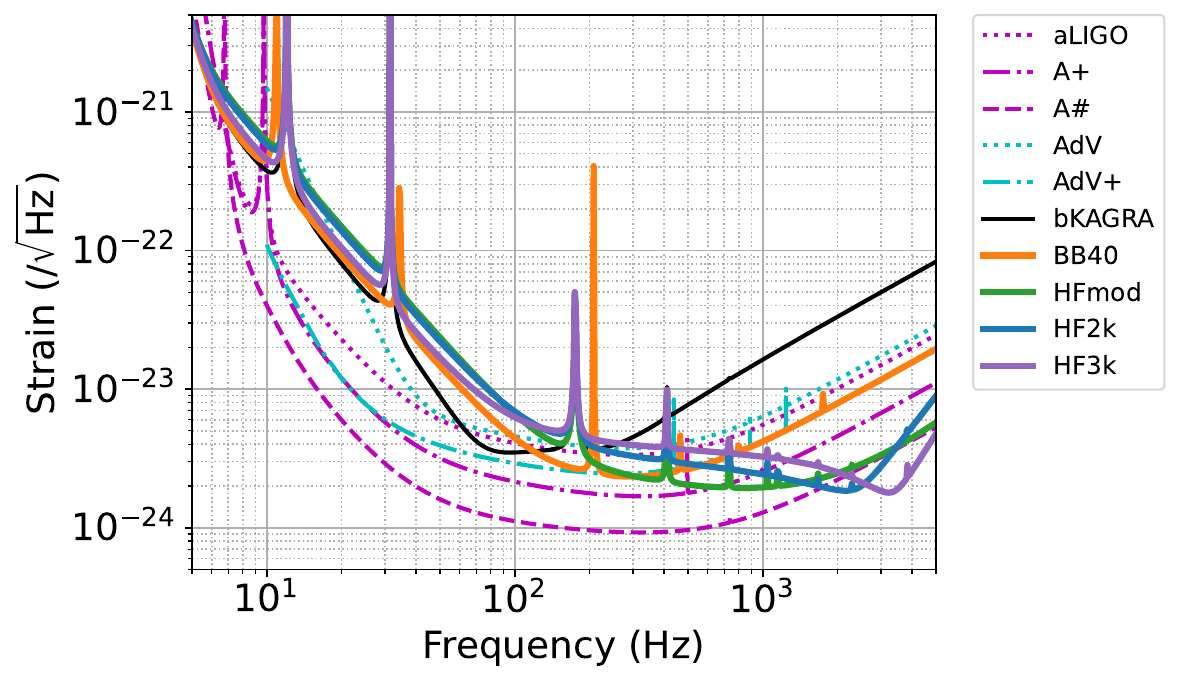}
	\caption{Strain sensitivity curves of the four KAGRA upgrade scenarios considered in this study (BB40, HFmod, HF2k, and HF3k). For comparison, the sensitivity curves of baseline KAGRA (bKAGRA), Advanced LIGO (aLIGO), A+, A\#, Advanced Virgo (AdV), and AdV+ are also shown. The KAGRA sensitivity curves are taken from Refs.~\cite{bKAGRACurve,KAGRA10yrCurve}, the LIGO sensitivity curves from Refs.~\cite{aLIGOCurve,AplusCurve,AsharpCurve}, and the Virgo sensitivity curves from Ref.~\cite{VirgoCurve}.} \label{fig:sensitivity}
\end{figure}

In the LIGO–Virgo–KAGRA collaboration, coordinated observing runs are essential, and the schedule of observational periods and detector upgrades is determined through collaboration among the three observatories~\cite{ObservingScenarioPaper,LVKObservingPlan}. The fourth observing run (O4) concluded in November 2025. The next observing run (O5) is planned following further detector upgrades, and a subsequent run is envisaged in the 2030s. In preparation for future observing runs, the LIGO detectors are pursuing a staged upgrade program toward A+ and A\#, while the Virgo detector is following a sequence of planned upgrades, including AdV+ and Virgo\_nEXT. In addition, the LIGO–India detector is under construction and is expected to join the network in the 2030s, further enhancing the international observatory network~\cite{LIGOIndia}.

In this context, a major upgrade of KAGRA after O5 is being considered, with a range of possible configurations presented in a KAGRA collaboration white paper~\cite{KAGRA10year}. In this study, we focus on four representative scenarios from the white paper that are expected to be achievable within the coming decade and that span distinct directions in detector sensitivity, as shown in Fig.~\ref{fig:sensitivity}.

These include a broadband sensitivity improvement scenario using heavier 40 kg sapphire test masses (BB40), as well as high-frequency optimized configurations that are designed to reshape the quantum noise spectrum in the kilohertz band. Among them, the HFmod scenario provides a global shift of the sensitivity toward higher frequencies, while the HF2k and HF3k options introduce localized sensitivity dips around 2 kHz and 3 kHz, respectively. These features are achieved by increasing the reflectivity of the signal recycling mirror, thereby entering the long signal recycling cavity regime~\cite{Miao2014,Somiya2020}.

All upgrade scenarios assume frequency-independent squeezing with an input squeezing level of 10~dB, together with high-quality-factor suspensions consistent with the baseline KAGRA (bKAGRA) design~\cite{bKAGRACurve,PSOKAGRA}. The bKAGRA design adopts a detuned resonant sideband extraction configuration to maximize the binary neutron star inspiral range, resulting in enhanced sensitivity at a narrow frequency band around 100~Hz. The BB40 configuration appears slightly shifted toward higher frequencies in Fig.~\ref{fig:sensitivity}, because frequency-independent squeezing improves the high frequency sensitivity at the expense of increased radiation pressure noise at low frequencies. Detailed detector parameters are summarized in Ref.~\cite{KAGRA10year,KAGRA10yearHW}, and are based on a series of technical studies within the KAGRA collaboration~\cite{KAGRAplus,KAGRAISWP2024}.

\section{Framework for evaluating localization performance} \label{sec:Framework}
The contribution of a detector upgrade to multimessenger observations cannot be assessed solely from the binary neutron star detection range, or equivalently the number of detectable binary neutron star mergers. Electromagnetic follow-up observations require sufficiently accurate localization in both angular and distance space, which depends on the combined performance of the detector network rather than on the sensitivity of an individual detector. Therefore, evaluating the impact of future detector upgrades on multimessenger observations requires a network-based assessment of localization performance.

As in previous studies~\cite{Kiendrebeogo2023,Alvin2026}, localization performance can be evaluated by injecting a large number of signals drawn from an assumed binary population into realizations of detector noise and performing full Bayesian parameter estimation for each event. However, the objective of this work is not to predict the localization performance for a realistic population, but rather to compare the relative localization performance of different KAGRA upgrade scenarios. We therefore employ the Fisher information matrix formalism, following the method used in Refs.~\cite{Hakeda2018,PSOKAGRA}. Throughout this paper, we quote sky localization areas $\Delta \Omega_{\rm s}$ and luminosity distance uncertainties $\Delta D_{\rm L}$ corresponding to the 90\% confidence region, obtained from the Fisher covariance matrix under the assumption of a Gaussian posterior distribution. The corresponding localization volume is then defined as a derived figure of merit,
\begin{equation}
\Delta V = D_{\rm L}^2 \, \Delta \Omega_{\rm s} \, \Delta D_{\rm L}.
\end{equation}

\begin{table}[t]
\begin{center}
  \caption{Source parameters used in the Fisher analysis. A GW170817-like binary neutron star system is placed at $z=0.03$, corresponding to a luminosity distance of 135~Mpc. Sky localization uncertainties are calculated for 1944 uniformly distributed combinations of $\{ \theta_{\rm s}, \phi_{\rm s}, \psi_{\rm p} \}$.} \label{tab:source_parameters}
\begin{tabular}{llcc}
 & Value  \\
\hline
Chirp mass & $\chirpmass=1.188 \Msun$ \\
Symmetric mass ratio & $\eta=0.248$ \\
Luminosity distance & $D_{\rm L} = 135$~Mpc \\
Inclination angle & $\iota=28^{\circ}$ \\
Colatitude & $\theta_{\rm s}$ \\
Longitude  & $\phi_{\rm s}$ \\
Polarization angle & $\psi_{\rm p}$ \\
Symmetric spin & $\chi_{\rm s}=0^{\circ}$ \\
Asymmetric spin & $\chi_{\rm a}=0^{\circ}$ \\
\end{tabular}
\end{center}
\end{table}

The gravitational-wave signal is modeled using the IMRPhenomD waveform, including amplitude and phase corrections up to 3.0 and 3.5 post-Newtonian order, respectively, as compiled in Ref.~\cite{IMRPhenomD}. As a representative source, we adopt a GW170817-like binary neutron star system placed at a redshift of $z=0.03$, corresponding to a luminosity distance of approximately 135~Mpc. The source parameters are summarized in Table~\ref{tab:source_parameters}. Because the localization uncertainty depends strongly on the source sky position and polarization angle, we calculate the localization uncertainty for 1944 uniformly distributed combinations of these parameters. Within the Fisher analysis in the low-redshift regime and for a fixed detector network, the sky localization area scales as $\Delta \Omega_{\rm s} \propto D_{\rm L}^2$, while the distance uncertainty scales as $\Delta D_{\rm L} \propto D_{\rm L}$. Approximate localization uncertainties for sources at different distances can thus be estimated by simple rescaling.

We assume a detector network consisting of the two LIGO detectors at Hanford and Livingston, the Virgo detector, and the KAGRA detector. From the sensitivity curves shown in Fig.~\ref{fig:sensitivity}, we adopt A\# for LIGO, AdV+ for Virgo, and one of the KAGRA upgrade scenarios for KAGRA. Since gravitational-wave detectors do not operate continuously, the localization performance of the network depends on the detector duty factors. To account for this effect, we repeat the localization analysis for all possible detector subnetworks and combine the resulting localization distributions according to the probability that each subnetwork is operational.

Let $p_S(\Delta \Omega_{\rm s})$, $p_S(\Delta D_{\rm L})$ and $p_S(\Delta V)$  denote the probability distributions of the sky localization uncertainty, luminosity distance uncertainty, and derived localization volume, respectively, for a detector subnetwork $S$. For simplicity, we explicitly show the construction for $\Delta \Omega_{\rm s}$, and an identical procedure is applied to $\Delta D_{\rm L}$ and $\Delta V$. The effective distribution of the network is given by
\begin{equation}
p_{\rm net}(\Delta \Omega_{\rm s}) = \sum_S w_S p_S(\Delta \Omega_{\rm s}),
\end{equation}
where $w_S$ is the probability that the detector subnetwork $S$ is operational,
\begin{equation}
w_S = \prod_{i\in S} d_i \prod_{j\notin S}(1-d_j),
\end{equation}
with $d_i$ denoting the duty factor of detector $i \in \{\mathrm{H}, \mathrm{L}, \mathrm{V}, \mathrm{K}\}$. The resulting network-averaged distribution of the sky localization uncertainty is used in Sec.~\ref{sec:Results} to evaluate the multimessenger performance of the different KAGRA upgrade scenarios. In Sec.~\ref{sec:Results}, we adopt a fiducial duty factor of 80\% for all detectors, representing a plausible operating condition for future detector networks.

To connect the localization performance with multimessenger observability, we introduce a threshold $\Delta \Omega_{\rm th}$ or $\Delta V_{\rm th}$ on the localization uncertainty. For a given luminosity distance $D_{\rm L}$, we evaluate the probability that the sky localization uncertainty satisfies $\Delta \Omega_{\rm s} < \Delta \Omega_{\rm th}$, denoted as $P_{\rm net}(\Delta \Omega_{\rm s} < \Delta \Omega_{\rm th} | D_{\rm L})$. This probability is obtained from the network-averaged sky localization distribution $p_{\rm net}(\Delta \Omega_{\rm s})$ defined above.

The expected number of well-localized events is then estimated by combining this probability with the binary neutron star merger rate density. We assume a constant comoving merger rate density $R_{\rm BNS} = 105.5~{\rm Gpc}^{-3}{\rm yr}^{-1}$, which is consistent with GWTC-3 estimates~\cite{GWTC3population} and previous KAGRA upgrade studies discussed in Ref.~\cite{KAGRA10year}. The corresponding detection rate below a given threshold is given by
\begin{equation}
\mathcal{R}(\Delta \Omega_{\rm th}) = \int_{0}^{D_{\rm L}^{\rm max}} dD_{\rm L} R_{\rm BNS} 4\pi D_{\rm L}^2\ P_{\rm net}(\Delta \Omega_{\rm s} < \Delta \Omega_{\rm th} | D_{\rm L}), \label{eq:rate}
\end{equation}
where we have used $dV = 4\pi D_{\rm L}^2 dD_{\rm L}$ in the low-redshift regime appropriate for the distance range considered in this work. The same procedure is also applied to $\Delta V$.

The upper limit $D_{\rm L}^{\rm max}$ is taken to be the smallest inspiral range among the detectors in the network, restricting the analysis to the high signal-to-noise regime where the Fisher approximation is expected to be reliable. The inspiral range is defined as the sky-averaged distance at which an equal-mass $1.4$--$1.4\Msun$ binary neutron star system yields a signal-to-noise ratio of $\rho_{\rm th} = 8$, following the standard convention~\cite{Creighton2011}. This calculation assumes that the binary neutron star population consists entirely of GW170817-like systems in evaluating the distance-dependent sky localization probability. While the true population is expected to be more diverse, this provides a representative fiducial model for comparing different detector configurations.

Overall, this framework provides a consistent and computationally efficient approach to evaluate and compare the multimessenger performance of different KAGRA upgrade scenarios under realistic detector network configurations.

\section{Multimessenger performance of KAGRA upgrade scenarios} \label{sec:Results}
Using the framework described in Sec.~\ref{sec:Framework}, we evaluate the multimessenger performance of the KAGRA upgrade scenarios introduced in Sec.~\ref{sec:Upgrade}. We first examine the localization performance of different detector networks, and then translate the results into the expected rate of well-localized binary neutron star mergers.

\begin{figure}[t]
	\centering
  \includegraphics[width=0.49\textwidth]{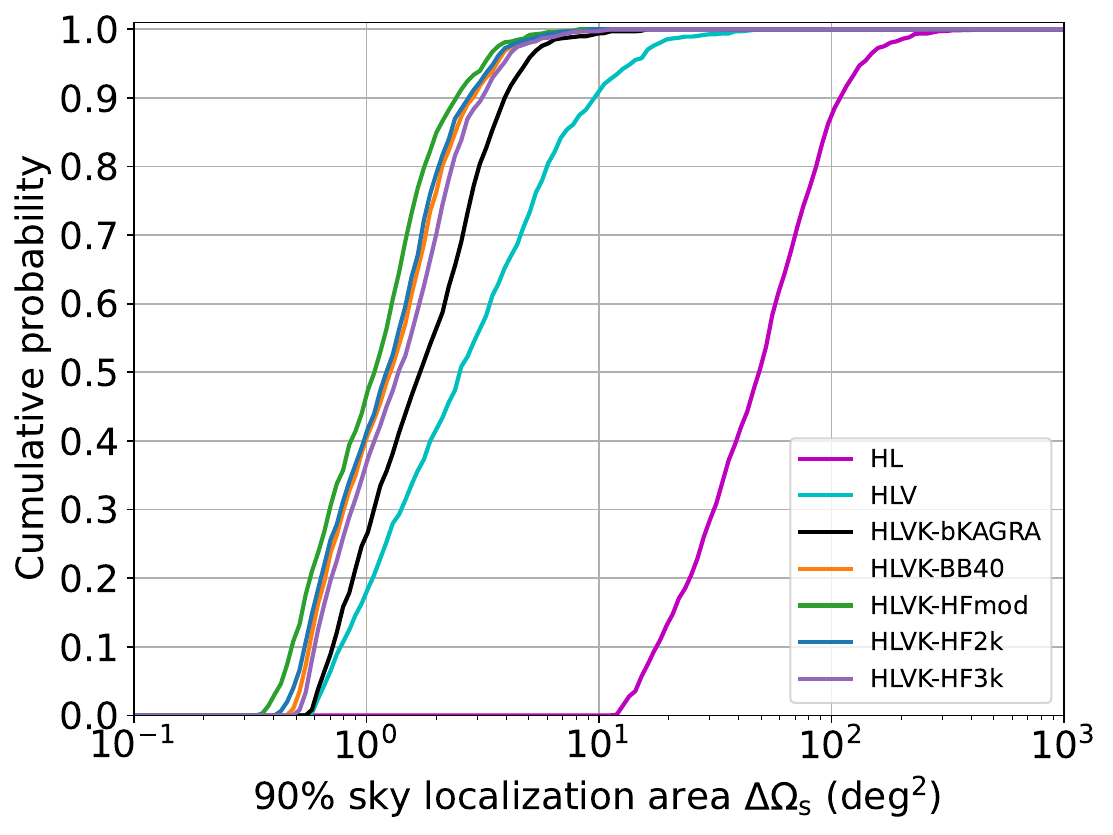}
  \hfill
  \includegraphics[width=0.49\textwidth]{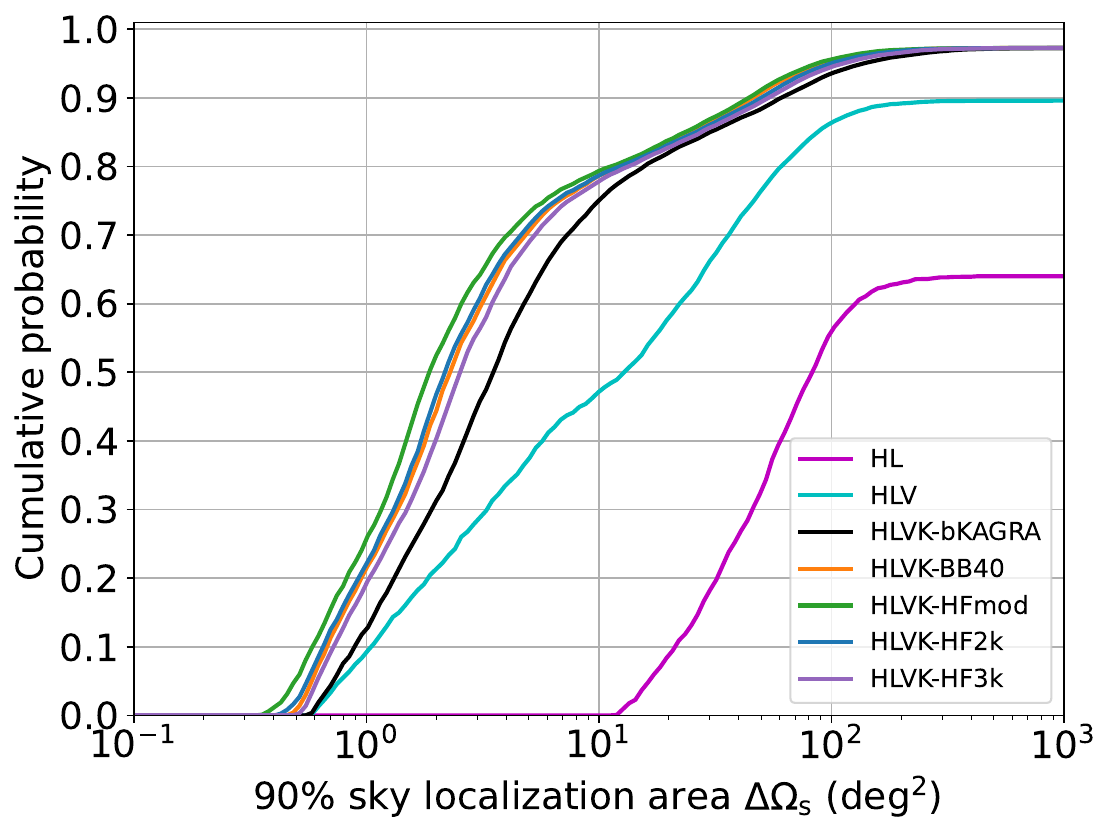} \\
    \caption{Cumulative distributions of sky localization areas for the binary neutron star sources described in Table~\ref{tab:source_parameters}, placed at a luminosity distance of 135~Mpc, observed with the HL, HLV, and HLVK detector networks. The left and right panels assume detector duty factors of 100\% and 80\%, respectively. The addition of Virgo and KAGRA significantly improves the sky localization performance. The impact of KAGRA becomes more apparent for realistic detector duty factors.} \label{fig:distribution}
\end{figure}

\begin{figure}[t]
	\centering
  \includegraphics[width=0.49\textwidth]{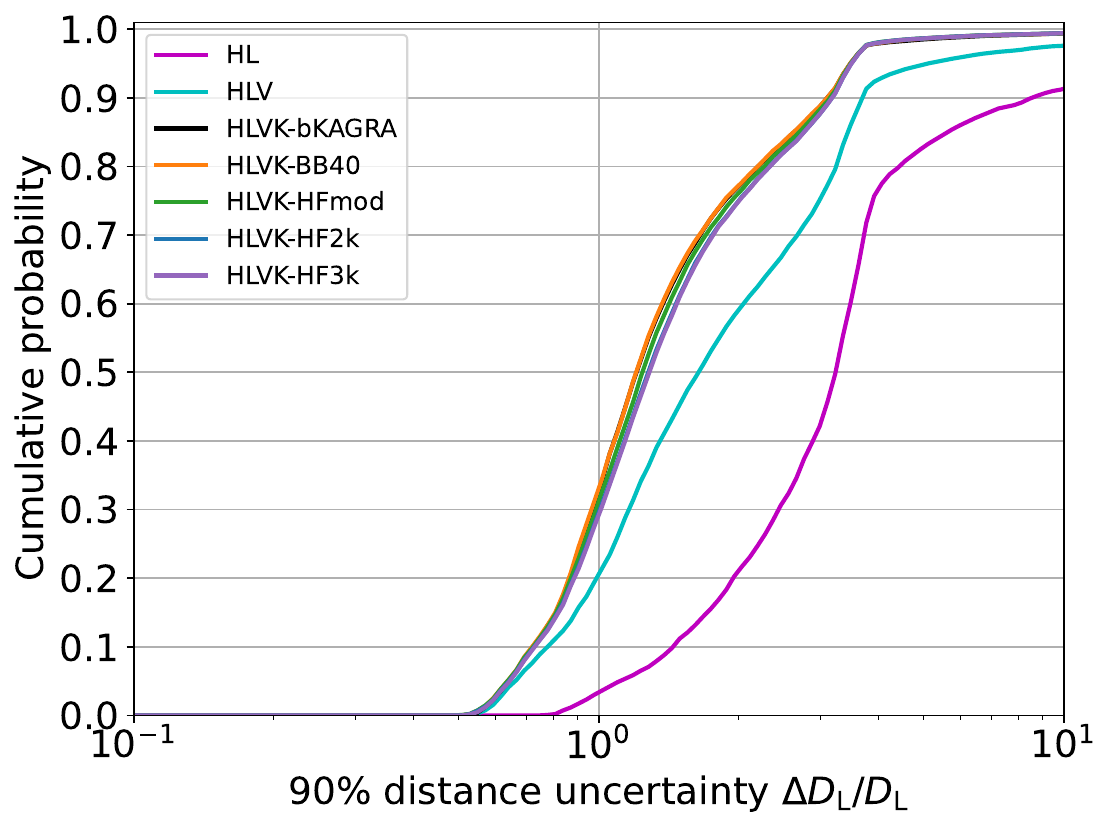}
  \hfill
  \includegraphics[width=0.49\textwidth]{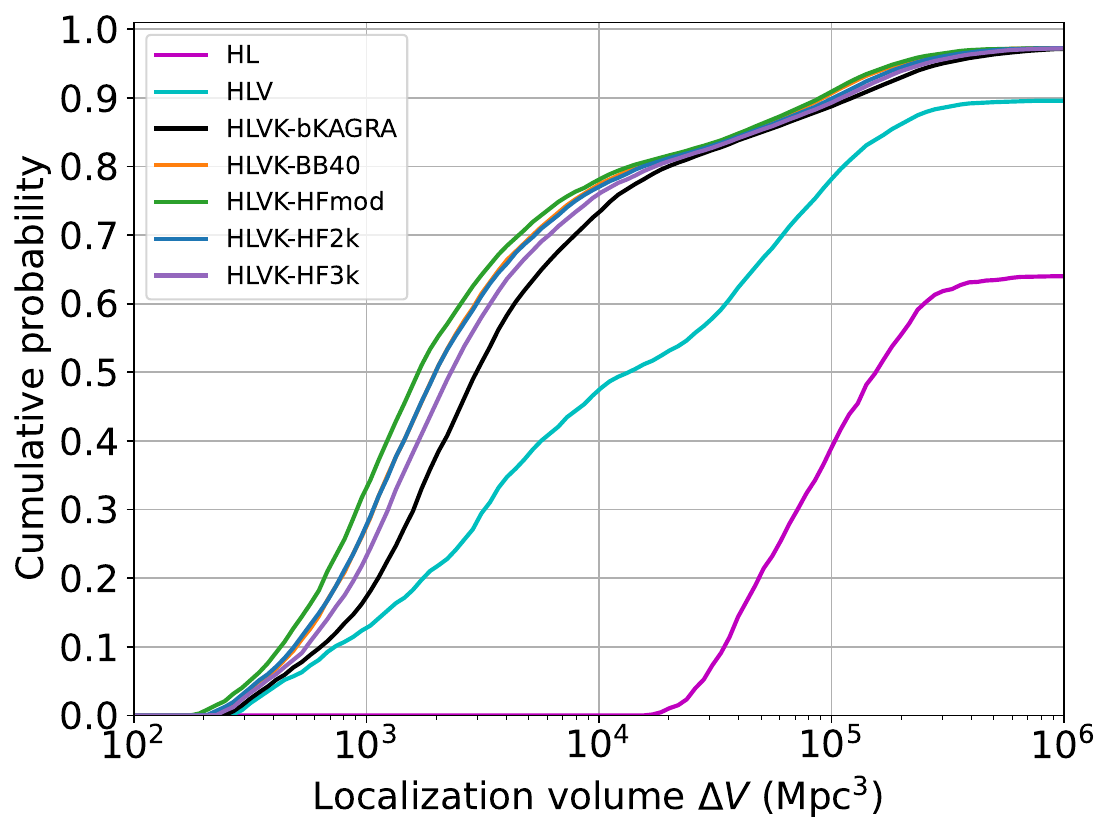}
    \caption{Same as Fig.~\ref{fig:distribution}, but for the luminosity distance uncertainties (left) and localization volume (right), assuming detector duty factors of 80\%.} \label{fig:distribution2}
\end{figure}

Figure~\ref{fig:distribution} shows the cumulative distributions of sky localization areas for the binary neutron star source described in Table~\ref{tab:source_parameters}, placed at a luminosity distance of 135~Mpc, for the different detector network configurations. The left panel assumes an idealized duty factor of 100\% for all detectors, while the right panel adopts a more realistic duty factor of 80\%. In the latter case, the cumulative probability does not reach unity because a fraction of events are observed with less than one detector and therefore do not yield meaningful sky localization constraints.

For a duty factor of 100\%, the addition of Virgo substantially improves the sky localization performance compared to the HL network, while the inclusion of KAGRA provides a further improvement. When a realistic duty factor is taken into account, the contribution of KAGRA becomes significantly more pronounced. For example, the probability of achieving a sky localization area smaller than $10~{\rm deg}^2$ increases from 47\% for the HLV network to more than 78\% for HLVK, depending on the KAGRA upgrade scenario. This behavior reflects the increased importance of an additional geographically separated detector when individual detectors are unavailable for a fraction of the observing time.

Similarly, Figure~\ref{fig:distribution2} shows the cumulative distributions of distance uncertainties and localization volumes, assuming detector duty factors of 80\%. The inclusion of Virgo and KAGRA enables polarization measurements, which break the degeneracy between luminosity distance and binary inclination, thereby improving distance estimation. Together with the improved sky localization, this leads to a reduction in the localization volume.

\begin{figure}[t]
	\centering
    \includegraphics[width=0.8\linewidth]{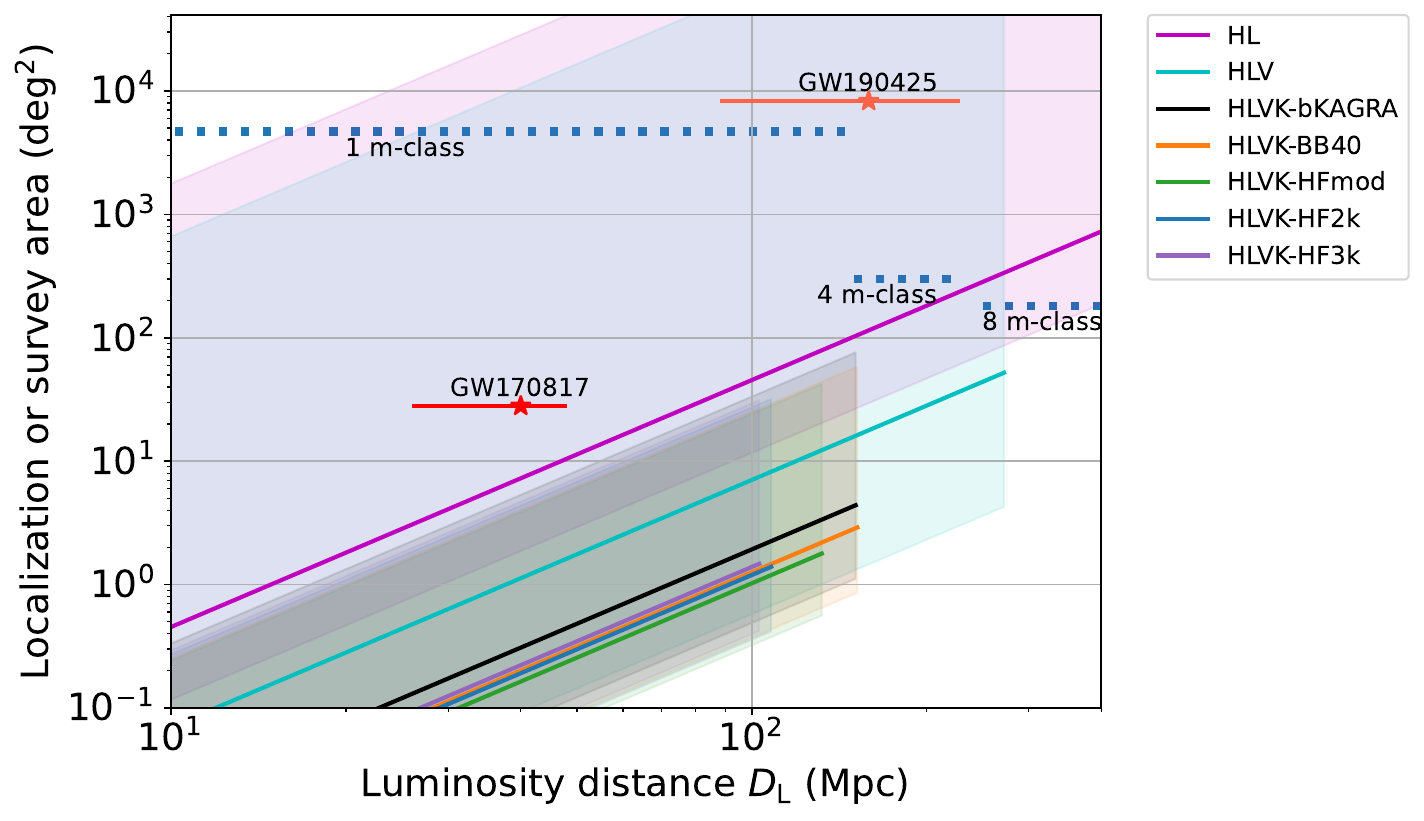} \\
    \includegraphics[width=0.8\linewidth]{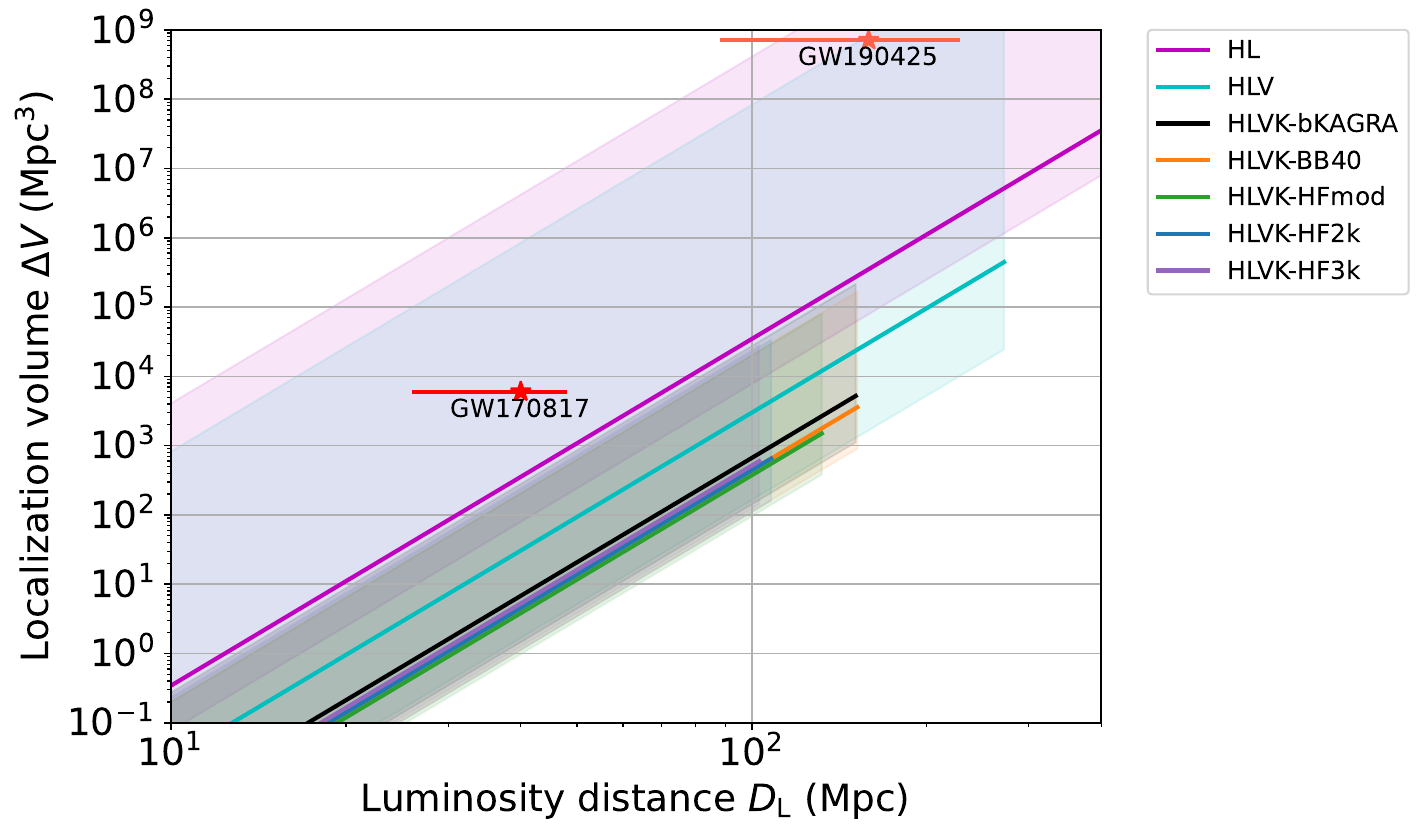}
    \caption{Sky localization area (top) and volume (bottom) as a function of luminosity distance for the binary neutron star source described in Table~\ref{tab:source_parameters}, assuming a detector duty factor of 80\%. Solid curves show the median localization for each detector network, while shaded regions indicate the 10th--90th percentile range. The curves are shown up to $D_{\rm L}^{\rm max}$, defined as the smallest inspiral range among the detectors in each network. The localizations of GW170817~\cite{GW170817} and GW190425~\cite{GW190425} are shown for reference. To illustrate the feasibility of electromagnetic follow-up observations, dotted lines in the top panel indicate survey areas corresponding to 100 times the field of view of wide-field 1~m-, 4~m-, and 8~m-class optical telescopes. The representative survey areas approximately match with the area covered by the actual follow-up observations so far (ZTF~\cite{ZTF_GW1,ZTF_GW2,ZTF_GW3}, DECam~\cite{DECam_GW1,DECam_GW2}, and Subaru/HSC~\cite{HSC_GW}) as well as the planned Target of Opportunity (ToO) survey area with Rubin/LSST~\cite{Rubin_ToO}.} \label{fig:dependence}
\end{figure}

The differences among the KAGRA upgrade scenarios are relatively modest compared to the differences arising from the detector network configuration itself. For the fiducial source at a fixed distance, HFmod yields slightly better localization area and volume than the other upgrade scenarios, owing to its improved sensitivity in the high-frequency inspiral phase~\cite{PSOKAGRA,KAGRAplus}. For luminosity distance measurements, BB40 performs best due to largest inspiral range.

The distance dependence of the localization performance is shown in Fig.~\ref{fig:dependence}. The localization areas and volumes at different distances are obtained by scaling the Fisher results at 135~Mpc according to the inverse dependence of the signal-to-noise ratio on luminosity distance. Among the KAGRA upgrade scenarios, HFmod provides the smallest localization areas at a fixed distance. In contrast, BB40 achieves the largest inspiral range and therefore remains sensitive to the most distant events.

The horizontal dotted lines for the localization area plot provide a rough indication of the sky areas that can be covered by typical optical and infrared follow-up facilities. For reference, the 1.2~m Zwicky Transient Facility (ZTF) has a field of view of approximately $47~{\rm deg}^2$ per pointing~\cite{ZTF}, the 1.8~m Pan-STARRS1 approximately $7~{\rm deg}^2$~\cite{PS1}, 4~m Blanco telescope equipped with the Dark Energy Camera (DECam) approximately 3 deg$^2$ ~\cite{DECam}, the 8.4~m Vera C. Rubin Observatory approximately $9.6~{\rm deg}^2$~\cite{Rubin}, and the 8.2~m Subaru Telescope equipped with Hyper Suprime-Cam (HSC) approximately $1.8~{\rm deg}^2$~\cite{HSC}. In general, larger-aperture telescopes can observe fainter counterparts and therefore probe events at greater distances, but this advantage comes at the cost of a smaller field of view. As a result, successful follow-up observations with 8~m-class facilities require substantially more accurate sky localization than those with 1~m-class facilities.

From this perspective, the localization performance of the HL network alone is often insufficient to fully exploit the reach of large-aperture follow-up facilities. Although the two LIGO detectors are sensitive to distant binary neutron star mergers, the resulting localization areas mostly exceed the survey areas achievable by 8\,m-class telescopes. Consequently, a significant fraction of distant events could remain challenging targets for efficient electromagnetic counterpart searches.

The addition of Virgo significantly improves the situation, bringing a substantial fraction of events within the survey areas achievable by 4~m- and even 8~m-class facilities. The inclusion of KAGRA further enhances the localization performance. Although KAGRA has a shorter inspiral range than the LIGO and Virgo detectors, its inclusion in the HLVK network substantially increases the fraction of nearby events localized within the survey areas of 1~m- and 4~m-class facilities, making successful electromagnetic counterpart identification considerably more robust.

A similar trend is observed for the localization volume. Assuming a typical galaxy number density of $(2$--$6)\times 10^{-3}$~Mpc$^{-3}$~\cite{Gehrels2016,Sasada2021}, a localization volume of $10^4$~Mpc$^3$ corresponds to approximately 20--60 potential host galaxies. With this number of galaxies, the expected number of contaminant transients, including unrelated supernovae, is small enough to perform detailed spectroscopic classification. Therefore, achieving a localization volume below $10^4$~Mpc$^3$ can significantly facilitate electromagnetic counterpart identification. Although this condition is primarily relevant for nearby events, the inclusion of KAGRA increases the probability of achieving localization volumes below $10^4$~Mpc$^3$, thereby facilitating host galaxy identification.

For reference, the measured localizations of GW170817~\cite{GW170817} and GW190425~\cite{GW190425} are also shown. Their locations in Fig.~\ref{fig:dependence} illustrate why extensive electromagnetic follow-up observations were successful for GW170817, while a counterpart search was more challenging for GW190425.

\begin{figure}[t]
	\centering
  \includegraphics[width=0.49\textwidth]{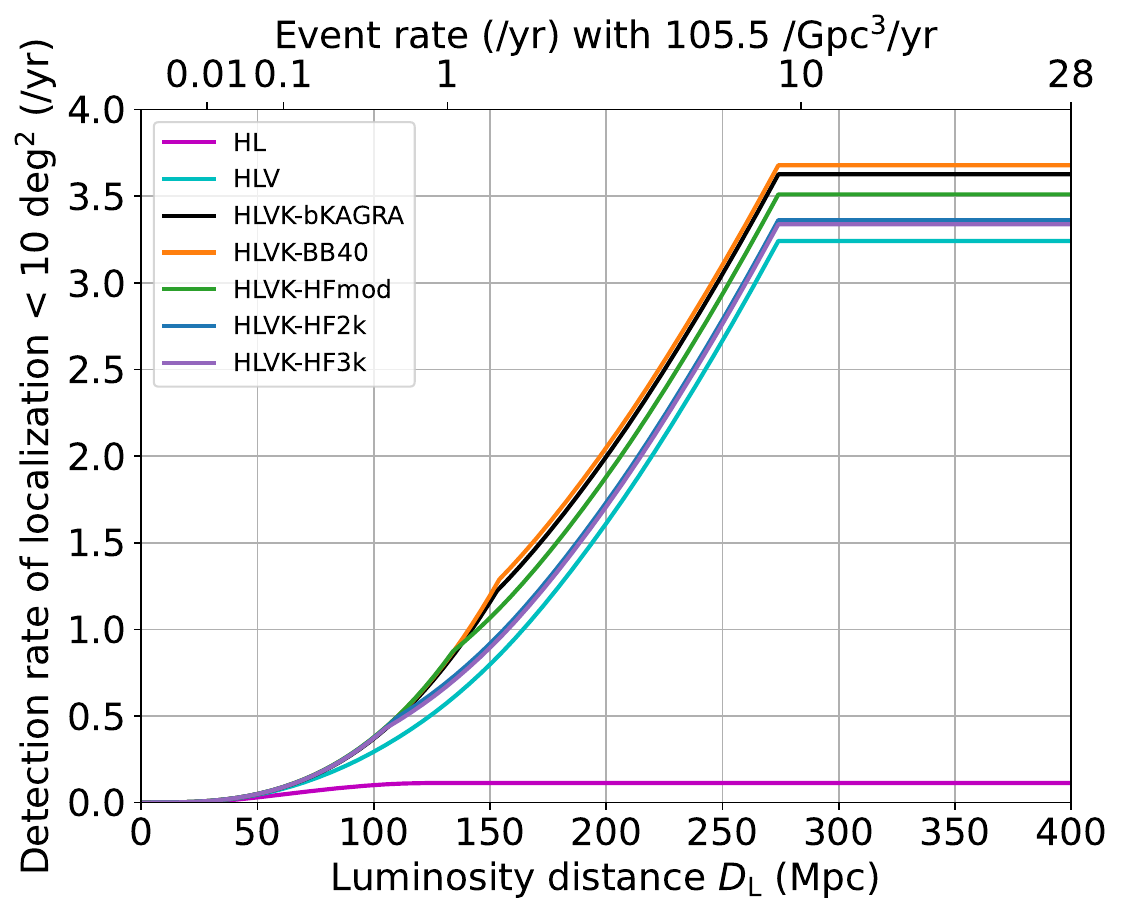}
  \hfill
  \includegraphics[width=0.49\textwidth]{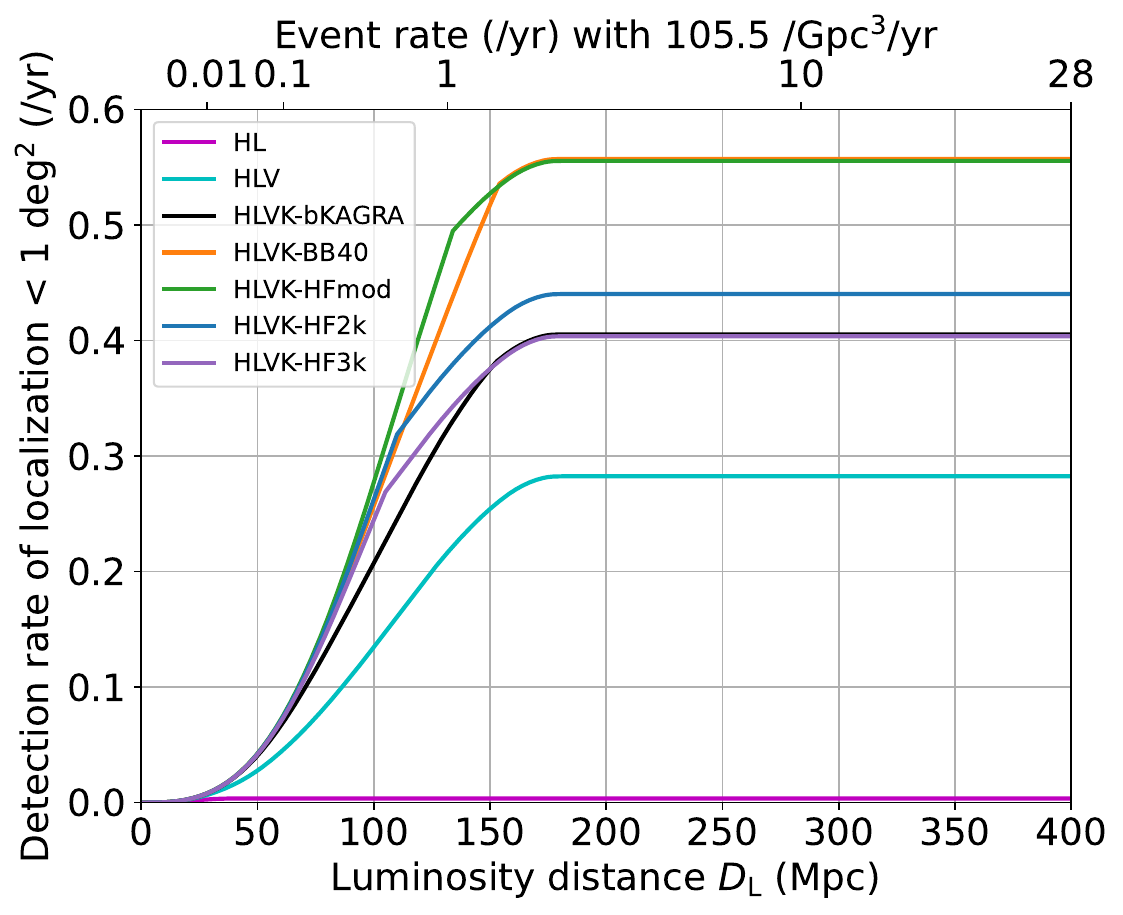} \\
  \includegraphics[width=0.49\textwidth]{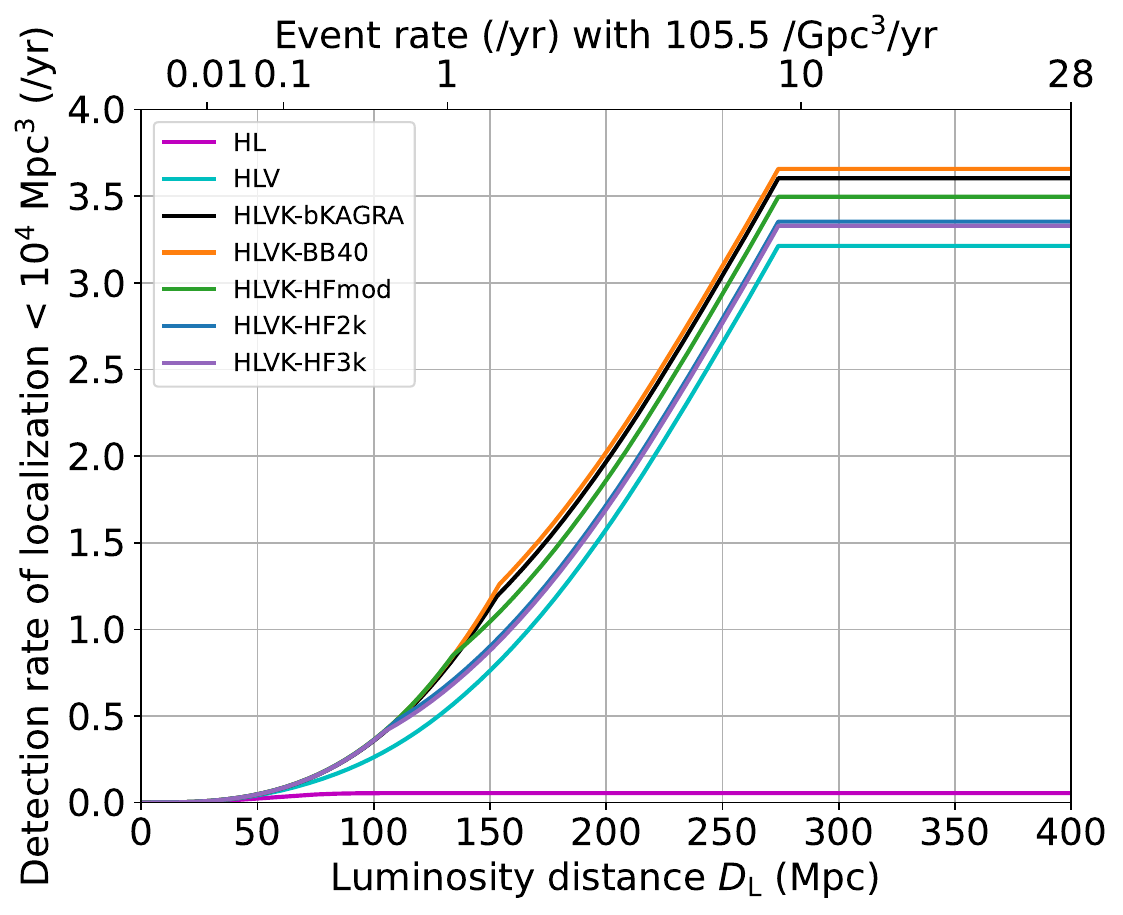}
  \hfill
  \includegraphics[width=0.49\textwidth]{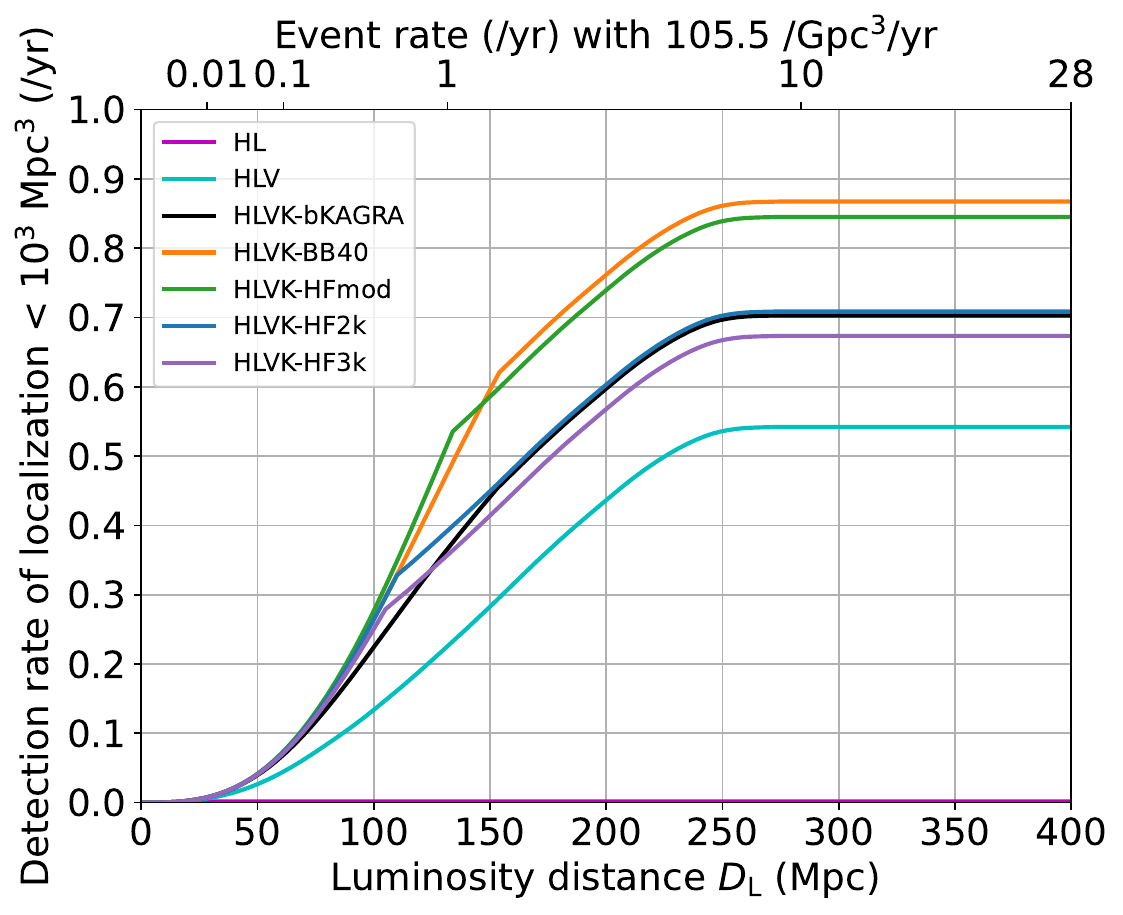}
    \caption{Expected cumulative annual rate of binary neutron star mergers localized within $10~{\rm deg}^2$ (top left), $1~{\rm deg}^2$ (top right), $10^4~{\rm Mpc}^3$ (bottom left) and $10^3~{\rm Mpc}^3$ (bottom right) as a function of luminosity distance, assuming a binary neutron star merger rate of $105.5~{\rm Gpc}^{-3}{\rm yr}^{-1}$ and a detector duty factor of 80\%. The detection rate at each distance is computed using Eq.~\ref{eq:rate}. When the luminosity distance exceeds the inspiral range of a detector, the detection rate is evaluated using the detector subnetwork that remains sensitive at that distance. The secondary x-axes show the cumulative event rate within a given luminosity distance.} \label{fig:detectionrate}
\end{figure}

The localization performance alone does not determine the multimessenger capability of a detector network, since the number of detectable events also depends on the detection range. We therefore combine the localization probability with the binary neutron star merger rate to estimate the expected rate of well-localized events.

Figure~\ref{fig:detectionrate} shows the cumulative annual rate of binary neutron star mergers that can be localized within $10~{\rm deg}^2$, $1~{\rm deg}^2$, $10^4~{\rm Mpc}^3$ and $10^3~{\rm Mpc}^3$ as a function of luminosity distance. At each distance, the event rate is obtained using Eq.~\ref{eq:rate}. Beyond the inspiral range of a given detector, the event rate is evaluated using the remaining operational subnetwork. For example, the HLVK curves are computed using the full HLVK network up to the KAGRA inspiral range, the HLV network between the KAGRA and Virgo inspiral ranges, and the HL network beyond the Virgo inspiral range.

The figure illustrates that the inclusion of Virgo leads to a substantial improvement in the rate of well-localized events compared to the HL network alone. For the $10~{\rm deg}^2$ or $10^4~{\rm Mpc}^3$ threshold, the addition of Virgo already captures most of the improvement, while KAGRA provides a modest further enhancement. For the more stringent $1~{\rm deg}^2$ or $10^3~{\rm Mpc}^3$ threshold, however, the contribution of KAGRA becomes more significant. In particular, for the BB40 configuration, the HLVK network yields an approximately 60\% higher rate of events with localization volumes below $10^3~{\rm Mpc}^3$ than the HLV network. At smaller luminosity distances, the HFmod configuration provides a better rate, reflecting improved sensitivity in the high-frequency band.

\begin{table}[t]
\begin{center}
  \caption{Summary of binary neutron star performance metrics for different detector networks. The binary neutron star inspiral range is listed in each column: the HL column corresponds to the LIGO A\# design sensitivity, the HLV column corresponds to AdV+, and the HLVK columns correspond to different KAGRA upgrade scenarios. Also shown are the median sky localization areas and volumes for the binary neutron star source specified in Table~\ref{tab:source_parameters}, at a luminosity distance of 135~Mpc, derived from Fig.~\ref{fig:distribution} and Fig.~\ref{fig:distribution2}. In addition, we show the annual detection rates of binary neutron star mergers localized within the thresholds, derived from Fig.~\ref{fig:detectionrate}. The best-performing KAGRA upgrade scenario for each metric is highlighted in bold.} \label{tab:summary}
\setlength{\tabcolsep}{4pt}
\begin{tabular}{lccccccc}
 & HL & HLV & bKAGRA & BB40 & HFmod & HF2k & HF3k \\
\hline
Range (Mpc) & 670 & 273 & 152 & {\bf 153} & 133 & 109 & 104 \\
Median $\Delta \Omega_{\rm s}$ (deg$^2$) & 82.7 & 12.8 & 3.51 & 2.29 & {\bf 1.87} & 2.18 & 2.53 \\
Median $\Delta V$ ($10^2$ Mpc$^3$) & 1552 & 134 & 30 & 20 & {\bf 17} & 20 & 23 \\
$< 10$ deg$^2$ (/yr) & 0.11 & 3.24 & 3.62 & {\bf 3.68} & 3.51 & 3.36 & 3.36 \\
$< 1$ deg$^2$ (/yr)  & 0.004 & 0.28 & 0.41 & {\bf 0.56} & 0.56 & 0.44 & 0.40 \\
$< 10^4$ Mpc$^3$ (/yr) & 0.06 & 3.21 & 3.60 & {\bf 3.66} & 3.50 & 3.35 & 3.33 \\
$< 10^3$ Mpc$^3$ (/yr) & 0.002 & 0.54 & 0.70 & {\bf 0.87} & 0.85 & 0.71 & 0.67 \\
\end{tabular}
\end{center}
\end{table}

The results discussed above are summarized in Table~\ref{tab:summary}. Among the KAGRA upgrade scenarios considered here, HFmod provides the best localization performance at a fixed distance. In contrast, BB40 yields the highest annual rate of well-localized binary neutron star mergers. Although bKAGRA and BB40 configurations have nearly identical inspiral ranges, BB40 achieves a higher rate of well-localized events. This difference arises because BB40 improves the detector sensitivity over a broader high-frequency band, whereas bKAGRA employs a detuned resonant sideband extraction configuration optimized for a narrower frequency region to maximize the inspiral range~\cite{PSOKAGRA}. These results indicate that, for both localization area and volume, broadband improvements in the high-frequency sensitivity are generally more beneficial than a narrow-band enhancement with a similar inspiral range.

The localization performance, however, represents only one aspect of the science case for future KAGRA upgrades. Previous studies have shown that HFmod provides the best performance in measuring the tidal deformability of neutron stars~\cite{KAGRA10year,KAGRA10yearScience}, although the extent to which this improvement translates into a meaningful gain in neutron star physics remains to be quantified. Similarly, HF3k has been found to maximize the expected annual number of detectable BNS post-merger signals. Nevertheless, even under optimistic assumptions, the expected detection rate remains below approximately $0.2~{\rm yr}^{-1}$~\cite{KAGRA10year,KAGRA10yearScience}.

The choice of a KAGRA upgrade scenario will ultimately depend on a broad range of scientific considerations. The results presented here indicate that, for binary neutron star multimessenger observations, the annual rate of well-localized events, both in terms of localization area and volume, is a useful figure of merit in addition to the inspiral range. The comparison results show that assessing the multimessenger impact of future detector upgrades requires consideration of both sky localization performance and detection range.

\section{Conclusions} \label{sec:Conclusions}

In this work, we developed a computationally efficient framework for evaluating the multimessenger performance of future gravitational-wave detector upgrades. The framework combines Fisher-matrix estimates of localization performance with detector duty factors and binary neutron star merger rates to estimate the expected number of well-localized events suitable for electromagnetic follow-up observations.

Using this framework, we compared several proposed KAGRA upgrade scenarios within the global LIGO--Virgo--KAGRA network. Our results show that assessing the multimessenger impact of future detector upgrades requires consideration of both localization performance and detection range. In particular, the upgrade scenario providing the best localization performance at a fixed distance is not necessarily the one that maximizes the annual rate of well-localized events. We also find that incorporating realistic detector duty factors increases the importance of network configuration, since a substantial fraction of events are observed by detector subnetworks rather than the full detector network. These results demonstrate the importance of evaluating detector upgrades using metrics directly connected to multimessenger observations, such as the expected rate of events localized within the survey area and volume of follow-up facilities.

Although this study focused on KAGRA upgrade scenarios, the framework itself is more generally applicable. By relying on a representative fiducial source and Fisher-based localization estimates, it enables rapid exploration of detector configurations and network assumptions while retaining the key features relevant for multimessenger observations. The method can therefore serve as a practical tool for future detector upgrade studies and network design investigations.


\ack
This work was supported by JSPS KAKENHI Grant Nos.~24K00640, 24K00649, 24K21546, 25K01008, 26H00407, 23H04894, 23H04891, 23H05432, 26K21726, 23H04893, and 26K21721, JSPS Birateral Program Grant Number JPJSBP120253204, JST FOREST Grant Numbers JPMJFR246G and JPMJFR212Y, JST NEXUS Grant Number JPMJNX26C7, and the Mitsubishi Foundation Research Grant Number 202510051.

The KAGRA project is supported by MEXT, JSPS Leading-edge Research Infrastructure Program, JSPS Grant-in-Aid for Specially Promoted Research 26000005, JSPS Grant-inAid for Scientific Research on Innovative Areas 2905: JP17H06358, JP17H06361 and JP17H06364, JSPS Core-to-Core Program A. Advanced Research Networks, JSPS Grantin-Aid for Scientific Research (S) 17H06133 and 20H05639 , JSPS Grant-in-Aid for Transformative Research Areas (A) 26A204: JP26H00407, the joint research program of the Institute for Cosmic Ray Research, University of Tokyo, National Research Foundation (NRF), Computing Infrastructure Project of Global Science experimental Data hub Center (GSDC) at KISTI, Korea Astronomy and Space Science Institute (KASI), and Ministry of Science and ICT (MSIT) in Korea, Academia Sinica (AS), AS Grid Center (ASGC) and the National Science and Technology Council (NSTC) in Taiwan under grants including the Science Vanguard Research Program, Advanced Technology Center (ATC) of NAOJ, and Mechanical Engineering Center of KEK. 

\section*{References}
\bibliographystyle{iopart_num}
\bibliography{KAGRAUpgradeChoice}
\end{document}